\documentstyle[preprint,prl,aps]{revtex}
\begin{document}
\draft
\title{\bf Glassy Solutions of the Kardar-Parisi-Zhang
Equation}

\author{ M.A. Moore, T. Blum, J.P. Doherty and M. Marsili}

\address{Department of Physics}
\address{University of Manchester}
\address{Manchester M13 9PL}
\address{United Kingdom}

\author{J-P. Bouchaud and P. Claudin}

\address{Service de Physique de l'Etat Condens\'e}
\address{Direction des Recherches sur l'Etat Condens\'e, les Atomes et les
Molecules}
\address{Commisariat \'a l'Energie Atomique, Orme des Merisiers}
\address{91191 Gif-sur-Yvette CEDEX, France}
\date{\today}
\maketitle

\begin{abstract}
It is shown that the mode-coupling equations for the strong-coupling
limit of the KPZ equation have a solution for $d>4$ such that the
dynamic exponent $z$ is $2$ (with possible logarithmic corrections)
and that there is a delta function term in the height correlation
function $\langle h({\bf k},\omega)h^*({\bf k},\omega)\rangle = (A/k^{d+4-z})
\delta (\omega/k^z)$ where the amplitude $A$ vanishes as $d \rightarrow 4$.
The delta function term implies that some features of the growing surface
$h({\bf x},t)$ will persist to all times, as in a glassy state.

\end{abstract}

\vfill


\newpage
\narrowtext

A simple non-linear Langevin equation has been proposed by Kardar, Parisi
and Zhang \cite{KPZ} and is now widely accepted as describing the
macroscopic properties of a wide variety of growth processes, such as
the Eden model, growth by ballistic deposition and the growth of an
interface in a random medium \cite{KS}.
\ This equation is also related to other seemingly disparate problems
such as randomly-stirred fluids \cite{FNS} (Burgers equation), dissipative
transport in the driven-diffusion equation \cite{BKS}, the directed
polymer problem in a random potential \cite{KZ} and the behavior of flux
lines in superconductors \cite{Hwa}.
\ Because of its ubiquity, any advance in understanding the KPZ equation is
likely to have wide significance in both the fields of nonequilibrium
dynamics and disordered systems.

The KPZ equation for a stochastically growing interface is:
\begin{equation}
{\partial h ({\bf x},t) \over \partial t} \ = \
\nu \nabla^2h \ + \ {\lambda \over 2} ({\bf \nabla h})^2 + \eta({\bf x},t).
\label{KPZ}
\end{equation}
It describes the large-distance, long-time dynamics of the growth
process specified by a single-valued height $h({\bf x},t)$ (i.e. one
with no overhangs or voids) on a $d$-dimensional substrate, ${\bf x} \in
R^d$.
\ This equation reflects the competition between the surface tension
smoothing forces, $\nu \nabla^2h$, the tendency for growth to occur
preferentially in the direction of the local normal to the surface,
represented by the nonlinear term in Eq. (\ref{KPZ}) and the Langevin noise
term $\eta$ which is added to model the stochastic nature of this growth
process.
\ The noise has zero mean and is Gaussian such that
\begin{equation}
\langle \eta({\bf x},t) \eta({\bf x^{\prime}},t^{\prime})\rangle \
=\ 2D \delta^d({\bf x}-{\bf x^{\prime}})\delta (t-t^{\prime}),
\label{noise}
\end{equation}
where $D$ specifies the noise amplitude.

The objective is to characterize the form of the surface.
\ Commonly studied are the correlation function
\begin{equation}
C({\bf k}, \omega) \ =\ \langle h({\bf k},\omega) h^*({\bf k},
\omega) \rangle
\label{correlation}
\end{equation}
and the response function:
\begin{equation}
G({\bf k},\omega) \ = \ {1 \over \delta^d({\bf k} - {\bf k^{\prime}}
) \delta(\omega-\omega^{\prime})} \biggl\langle {\partial h({\bf k},
\omega) \over \partial \eta ({\bf k^{\prime}}, \omega^{\prime})}
\biggr\rangle .
\label{response}
\end{equation}

The correlation and response function take the scaling forms:
\begin{equation}
C({\bf k},\omega) \ = \ {1 \over k^{2\chi +d+z}}\ n\left( {\omega
\over k^z} \right)
\label{corr-scale}
\end{equation}
\begin{equation}
G({\bf k},\omega) \ = \ {1 \over k^{z}}\ g \left( {\omega
\over k^z} \right) .
\label{response-scale}
\end{equation}
For $d>2$, there are two distinct regimes.
\ There is a weak-coupling regime, for $\lambda < \lambda_c$, where
the nonlinear term is irrelevant and $z=2$ and $\chi=(2-d)/2$.
\ For $\lambda>\lambda_c$, the non-linear term is relevant and the
scaling relation $\chi+z=2$ follows from the invariance of Eq. (\ref{KPZ})
to an infinitesimal tilting of the surface $h \rightarrow h+
{\bf v}\cdot {\bf x}$, ${\bf x} \rightarrow {\bf x}-\lambda {\bf v}
t$ \cite{FNS}.
\ There is thus only one independent exponent to be determined
in the strong-coupling regime, (which is the only one we shall
consider here).

Because there are no obvious small parameters to describe the
strong-coupling regime, most studies of it have been numerical
\cite{FT1}.
\ The most recent numerically determined values of the dynamic
exponent $z$ seem to lie between the Wolf-Kertesz \cite{WK} conjecture
$\chi/z = 1/(2d+1)$ and that of Kim and Kosterlitz \cite{KK}
$\chi/z=1/(d+2)$ when $d<4$.
\ For $d \geq 5$, $\chi/z$ is still apparently non-zero but lies below
the values predicted by both conjectures.
\ Thus there is some very weak numerical evidence that the upper
critical dimension $d_c$ beyond which $z=2$ and $\chi=0$ is $4$.
\ In our studies we find $d_c=4$ and we believe that for $d>4$ the
apparent non-zero values of $\chi$ arise from logarithmic factors
masquerading as small powers.

We take a non-perturbative approach to the strong-coupling regime
called mode-coupling theory \cite{BKS}; in it one retains in the
diagrammatic expansion for $C$ and $G$ only diagrams which do not
renormalize the three-point vertex $\lambda$.
\ This procedure leads to the coupled equations:
\begin{equation}
G^{-1}({\bf k},\omega) = G_0^{-1}({\bf k},\omega) +
\lambda^2 \int {d \Omega \over 2 \pi} \int {d^d q \over (2 \pi )^d}
[{\bf q}\cdot ({\bf k}-{\bf q})][{\bf q}\cdot{\bf k}] G({\bf k}-
{\bf q}, \omega -\Omega)C({\bf q},\Omega)
\label{mode1}
\end{equation}
\begin{equation}
C({\bf k},\omega) = C_0({\bf k},\omega) +
{\lambda^2 \over 2} |G({\bf k},\omega)|^2
\int {d \Omega \over 2 \pi} \int {d^d q \over (2 \pi )^d}
[{\bf q}\cdot ({\bf k}-{\bf q})]^2 C({\bf k}-
{\bf q}, \omega -\Omega)C({\bf q},\Omega),
\label{mode2}
\end{equation}
where $G_0({\bf k},\omega)=(\nu k^2 -i\omega)^{-1}$ is the bare response
function and $C_0({\bf k},\omega)=2D|G({\bf k},\omega)|^2$.
\ Some of us \cite{DMKB} have recently shown that the mode-coupling equations
arise from the large-$N$ limit of a generalization of the KPZ
equation to an $N$-component model.
\ In principle, this might allow a systematic expansion in $1/N$,
by which one could go systematically beyond mode-coupling equations
towards a solution to the full problem.

In the strong-coupling limit, the scaling functions $n(x)$ and
$g(x)$, where $x=\omega/k^z$, satisfy the equations
\begin{equation}
g^{-1}(x)\ = \ -ix + P_1 I_1(x),
\label{mode3}
\end{equation}
\begin{equation}
n(x) \ = \ {1 \over 2} P_1 |g(x)|^2 I_2(x),
\label{mode4}
\end{equation}
where $P_1=\lambda^2/(2^{d}\Gamma({d-1 \over 2}) \pi^{d+3/2})$.
The integrals $I_1$ and $I_2$ are given by \cite{Tu}
\begin{eqnarray}
I_1(x) &=& \int_0^{\pi} d\theta \sin^{d-2}\theta
\int_{-\infty}^{+\infty}dy \int_0^{\infty}dq \cos \theta (
\cos \theta -q)
\nonumber \\
&& \ \ \ \ \ \ \ \ \
\times {q^{2z-3} \over r^z}
g\Biggl({x-q^zy \over r^z} \Biggr) n(y),
\label{I1}
\end{eqnarray}
\begin{eqnarray}
I_2(x) &=& \int_0^{\pi} d\theta \sin^{d-2}\theta
\int_{-\infty}^{+\infty}dy \int_0^{\infty}dq (
\cos \theta -q)^2
\nonumber \\
&& \ \ \ \ \ \ \ \ \
\times {q^{2z-3} \over r^{\Delta}}
n\Biggl({x-q^zy \over r^z} \Biggr) n(y),
\label{I2}
\end{eqnarray}
where $r=\Bigl(1+q^2-2q\cos\theta \Bigr)^{1/2}$ and $\Delta =
d+4-z$.

In the strong-coupling limit, the bare term in $D$ in Eq. (\ref{mode1})
can be dropped, as can the term $\nu k^2$ in the bare propagator
$G_0$.
Eqs. (\ref{mode3}-\ref{I2}) are valid for the limit $\omega
\rightarrow 0$, $k^z \rightarrow 0$ with $\omega/k^z$ fixed.
\ Notice that provided $z<2$ no cutoff is needed.

The numerical solution of the mode-coupling equations in Eqs. (\ref{mode3}
-\ref{I2}) presents formidable problems.
\ Recently Tu \cite{Tu} has attempted such a numerical solution, but
the dependence which he obtained for $z$ on $d$ (first increasing from
the exact value of $z=3/2$ in $d=1$ then decreasing at larger $d$) is so
strange that we suspect that his solution cannot be accurate.
\ We suspect from our own attempts at finding a direct numerical
solution that problems can arise from the integrable singularities
in eqs. (\ref{I1}) and (\ref{I2}).
\ However, if we first assume that $d_c=4$ so that for $d>d_c$, $z=2$
and that $n(x)=A\delta(x)$ then progress is possible.
\ (We shall later confirm that such assumptions are consistent.)

If $z=2$, it is now no longer possible to drop $\nu k^2$ from
the bare propagator.
\ Furthermore, $I_1(x)$ ceases to be well-defined as the final momentum
integral diverges logarithmically without a cutoff $\Lambda$.
\ Thus even if $z=2$ for $d=4$, the scaling of $\omega$ is not likely
to be simply with $k^2$ but with $k^2$ modified by some (unknown to
us) power of $log(\Lambda/k)$.
\ We have been unable to make any analytic progress once cutoffs are
explicitly required.
\ Instead we shall study the following problem in which the bare
propagator is:
\begin{equation}
G_0^{-1}({\bf k},\omega) \ = \ \nu(z^*)k^{z^*} -i \omega,
\label{new-response}
\end{equation}
and we shall imagine that $z^*$ is arbitrarily close to $2$.
\ With $z^*<2$ no cutoff is required.
\ Setting $n(x)=A\delta(x)$, eq. (\ref{mode3}) then becomes:
\begin{equation}
g^{-1}(x) \ = \ g^{-1}(0)-ix + P_1\left(I_1(x)-I_1(0)\right).
\label{new-mode3}
\end{equation}
The integrals defining the difference $(I_1(x)-I_1(0))$ are convergent
without a cutoff and can be calculated with $z=2$.
\ Also $g^{-1}(0)=\nu(z^*)+I_1(0)$ where $I_1(0)= \lambda^2 A g(0)
T_1(z^*,d)$ with
\begin{equation}
T_1(z,d) ={ \left(3-2z \right)
\Gamma \left({d-z \over 2}+1 \right) B(1-{z\over 2},z-1)
 \over (4 \pi)^{d/2+1} \Gamma ({d \over 2} +2-z )
\Gamma \left({d+z \over  2} \right)},
\label{T1}
\end{equation}
where $B(x,y)$ is the beta function.
\ In addition, with $n(x)=A\delta(x)$, eq. (\ref{mode4}) becomes:
\begin{equation}
A \ = \ \lambda^2 A^2 |g(0)|^2 T_2(z^*,d),
\label{new-mode4}
\end{equation}
where
\begin{equation}
T_2(z,d) = { \Gamma \left({d \over 2}+2 -2z\right)
B(z,z-1) \over (4 \pi)^{d/2+1}
\left[ \Gamma \left({d\over 2} +2-z \right) \right]^2 }
\left[ {d \over 2} +2(z-1)(z-2) \right].
\label{T2}
\end{equation}
Notice that $T_1(z,d)$ is divergent as $z \rightarrow 2$ reflecting
the need for a cutoff in that limit.
\ The solution of (\ref{new-mode4}), setting $z^*=2$ (the integrals
here are convergent in that limit) is:
\begin{equation}
 A |g(0)|^2 \ = \  {(d-4)(d-2) \over P_1~d}
{4 ~\Gamma(d/2) \over \Gamma(1/2) ~\Gamma ((d-1)/2)}.
\label{A-eq}
\end{equation}
Setting $g(0)=1$ (as can always be achieved by adjusting $\nu(z^*)$,
the equation for $g(x)$ is:
\begin{eqnarray}
g^{-1}(x) \ &=& \ 1 -ix - B \int_0^1 dr ~r^{d-1}\Bigl(g\left(
{x \over r^2} \right)-1 \Bigr)
\nonumber \\
&&\ \ \
- B \int_1^{\infty} {dr \over r}
\Bigl( g \left({x \over r^2} \right) -1\Bigr) ,
\label{g-iter}
\end{eqnarray}
where $B=4(d-4)(d-2)/d^2$.
\ This equation is readily solved for $g(x)$ numerically.
\ Notice that as $d \rightarrow 4$, $g^{-1}(x) \rightarrow
1 - ix $ and $A \rightarrow 0$.

We have therefore found an exact solution for the mode-coupling equations
for $d > 4$ when the bare propagator is as given by equation
(\ref{new-response}) with $z^*<2$.
\ The model with this bare propagator (\ref{new-response})
is in some sense ``long-ranged" compared to the model with the
conventional bare propagator ({\it i.e.} with $z^*=2$)
\ If the value of $z$ emerging from the mode-coupling approach with
the new bare propagator (eq. (\ref{new-response})) had been {\it less}
than the bare $z^*$, that is, if the renormalized propagator were
even longer ranged, then the new model (with $z^*<2$) and the
conventional model (with $z^*=2$) would belong to the same
(strong-coupling) universality class.
\ However, this was not found.
\ The $z$ of the calculation is equal to the $z^*$ of the bare
propagator.
\ One concludes that the value of $z$ associated with the
conventional ``short-range" propagator must then be greater than
or equal to $z^*$.
\ But as $z^*$ can be taken arbitrarily close to $2$, we conclude
that the true value of $z$ associated with the true ``short-range"
propagator must be $2$, up to logarithmic factors.

One can check whether the solutions for $n(x)$ and $g(x)$ are iteratively
stable as follows.
\ By writing $n(x)=A\delta(x) + p_{n+1}(x)$ in Eq. (\ref{mode4}), etc.,
one sees
that:
\begin{eqnarray}
&&p_{n+1}(x) = B |g(x)|^2
\Biggl[ \int_0^1 {dq \over q} p_n\left({x \over q^2}\right)
\nonumber \\
&&\ \ + \int_1^{\infty}{dq \over q^{d+1}} \left[d~q^2 - (d-1) \right]
p_n\left({x \over q^2 } \right) \Biggr]
+ O(p_n^2) .
\label{corr-iter}
\end{eqnarray}
Under iteration we found that $p_{n+1}(x)\rightarrow \lambda_R ~p_n(x)$
as $n\rightarrow \infty$, with the eigenvalue $\lambda_R<1$ (which
implies stability) for $d< d^* < 4.76...$.
\ In fact there is a relation between the eigenvalue $\lambda_R$ and the
behavior of $p_n(x)$ as $x \rightarrow 0$; if $p_n(x) \rightarrow
D/x^a$ as $x \rightarrow 0$, then direct substitution into Eq.
(\ref{corr-iter}) shows that:
\begin{equation}
\lambda_R \ = \ {B \over 2}
\left[ {1 \over a} ~+~ {d \over d/2 -1 -a} ~-~
{d-1 \over d/2-a} \right].
\label{lamda}
\end{equation}
$\lambda_R$ has a minimum as a function of $a$.
\ Within our limited numerical accuracy, $\lambda_R$ determined by
iteration of eq. (\ref{corr-iter}) is equal to this minimum value.
\ When $\lambda_R>1$, i.e. when $d>d^*$ the simple delta function
solution is unstable.
\ We then expect that $n(x)=A \delta(x) + p(x)$ where $p(x)$ is proportional
to $(d-d^*)p_{\infty}(x)$ and $p_{\infty}(x)$ is the limiting form for
$p_n(x)$ as $n \rightarrow \infty$.

Thus we have a solution of the mode-coupling equations for
$d>4$, which is exact in the scaling limit and stable.
\ It is a ``glassy'' solution, in that on Fourier transforming  to
$({\bf k},t)$ variables, one sees that $C({\bf k},t)$ (in the
scaling limit) is constant in time.
\ This is rather like the original definition of Edwards and
Anderson \cite{EA} of spin-glass order, i.e., the spins $S_i(t)$ have such
order if $C(t)={1 \over N} \Sigma \langle S_i(0)S_i(t)\rangle
\ne 0$ as $t \rightarrow \infty$ so that the
Fourier transform of $C(t)$ has a $\delta$-function in it.
\ However, in the present case, quenched disorder is {\it a priori}
absent, as in a `true' glass.
\ Hence the KPZ equation may well be another interesting model where
quenched disorder is `self-generated', as recently proposed and discussed
in  \cite{glasses}.
\ If this scenario is correct, our implicit assumption that the
correlation and response functions are time translational invariant may
not be valid, and the mode coupling may have to be recast in a
two-time formulation. \cite{CuKu}.

One might wonder if the glassy behavior is attributable to the
approximations made in the mode-coupling equations.
\ While our solutions of $n(x)$ are only within the context of
mode-coupling, it is easy to see that non-mode-coupling diagrams
for $C$, (see Fig. 1) are such that if the $\delta$-function
ansatz is inserted for the correlator within the diagram, then
each of these diagrams gives only a $\delta$-function contribution
to $n(x)$.
\ Moreover, explicit evaluation of higher-order diagrams permits a
generalization of eq. (\ref{A-eq}) which as $d \rightarrow 4^+$
takes on the form:
\begin{equation}
A \ =\  C_2 \left({\lambda^2 A^2|g(0)|^2 \over d-4} \right)
+ C_3 \left({\lambda^4 A^3|g(0)|^4 \over (d-4)^2} \right)
+ \ldots,
\label{beyond}
\end{equation}
where $C_2$ and $C_3$ are constants.
\ Eq. (\ref{beyond}) implies that provided there is a nontrivial
solution, it will always be such that $\lambda^2 A |g(0)|^2
\sim (d-4)$.
\ Hence we expect the upper critical dimension $d_c$ to be $4$
even beyond the mode-coupling approximation.

An approximate solution of Eqs. (\ref{mode1}) and (\ref{mode2})
has also suggested that $d_c\approx 3.6$ \cite{DMKB} \cite{BC}.
\ Previously Bouchaud and Georges \cite{BG} had argued that
$d_c \geq 4$ based on a comparison with directed percolation.
\ The existence of a finite $d_c$ is supported by a $1/d$ expansion
\cite{CD}; in addition, a prediction that $d_c$ is $4$ is
contained in the functional $RG$ calculation of Halpin-Healy
\cite{HH}.

For dimensions $d<4$, we do not expect to see this $\delta$-function,
but precursors of glassy behavior such as very long-lived peaks
in $h({\bf x},t)$ are known to exist for $d=2$.
\cite{Tim}
\ It would be valuable to do numerical studies of the scaling limit
of $C({\bf k},\omega)$ for $d>4$ to check the existence of glassy
behavior.

Finally, we speculate that given the solution for $d=4$ in the
mode-coupling equations it should be possible to construct a
perturbative expansion for $z$ in $\epsilon$, where $\epsilon=
4-d$.
\ So far, however, we have not succeeded in this aim.

Two of us (MAM and TB) would like to thank the Newton Institute,
Cambridge for its hospitality during the performance of this work.


\bibliographystyle{unsrt}

\begin{figure}
\end{figure}

\begin{description}

\item{Fig. 1} Diagrams for the correlation function beyond the
mode-coupling approximation.
\ The lines with the circles within them are height correlation
functions.
\ If a $\delta$-function form is inserted for them, the diagrams themselves
give $\delta$-function contributions to $n(x)$.

\end{description}

\end{document}